\documentclass[prl,superscriptaddress,twocolumn]{revtex4}
\usepackage{epsfig}
\usepackage{latexsym}
\usepackage{amsmath}
\begin {document}
\title {Synchronization and partial synchronization of linear maps}
\author{Adam Lipowski}
\affiliation{Department of Physics, University of Geneva, CH 1211
Geneva 4, Switzerland}
\affiliation{Faculty of Physics, A.~Mickiewicz University,
61-614 Pozna\'{n}, Poland}
\author{Michel Droz}
\affiliation{Department of Physics, University of Geneva, CH 1211
Geneva 4, Switzerland}
 %%%%%%%%%%%%%%%%%%%%%%%%%%%%%%%%%%%%%%%%%%%%%%%%%%%%%%%%%%%%%%%%%%%%%%%%%%%%%%%
\begin {abstract}
We study synchronization of low-dimensional ($d=2,3,4$) chaotic piecewise linear maps.
For Bernoulli maps we find Lyapunov exponents and locate the synchronization transition, that numerically is found to be  discontinuous (despite continuously vanishing Lyapunov exponent(s)).
For tent maps, a limit of stability of the synchronized state is used to locate the synchronization transition that numerically is found to be continuous.
For nonidentical tent maps at the partial synchronization transition, the probability distribution of the synchronization error is shown to develop highly singular behavior.
We suggest that for nonidentical Bernoulli maps (and perhaps some other discontinuous maps) partial synchronization is merely a smooth crossover rather than a well defined transition.
More subtle analysis in the $d=4$ case locates the point where the synchronized state becomes stable.
In some cases, however, a riddled basin attractor appears, and synchronized and chaotic behaviors coexist.
We also suggest that similar riddling of a basin of attractor might take place in some extended systems where it is known as stable chaos.
\end{abstract}
\maketitle
%%%%%%%%%%%%%%%%%%%%%%%%%%%%%%%%%%%%%%%%%%%%%%%%%%%%%%%%%%%%%%%%%%%%%%%%%%%%
\section{Introduction}
Recently, synchronization of chaotic dynamical systems has been intensively studied~\cite{FUJISAKA}.
To some extent this is motivated by its numerous experimental 
realizations in lasers, electronic circuits or chemical 
reactions~\cite{EXP}.
Of interest, however, are also theoretical aspects of this phenomenon.
Relatively well understood is the problem of synchronization of identical systems.
In this case it is known that synchronization might appear only  when the so-called transversal, or conditional~\cite{PECORA} Lyapunov exponents are negative~\cite{FUJISAKA}.
Much less understood, however, remains the problem of partial synchronization that occurs when systems are nonidentical.
In such a case chaotic systems do not fully synchronize which causes fundamental problems even with the very detection of synchronization.
In addition to studying Lyapunov spectrum, some other methods to detect this transition were proposed. 
For example it was noticed that at the partial synchronization of  continuous dynamical systems driven by common noise, the probability 
distribution of phase difference changes~\cite{ZHOU}.
One of the difficulties in studying synchronization is the lack of analytical insight into this problem.
Consequently, most of the results in this field are based  on numerical calculations.
It would be desirable to study models that would be exactly solvable if not fully then at least with respect to some properties.

In the present paper we study low-dimensional ($d=2,3,4$) linear maps.
Such systems have the advantage that under some conditions their Lyapunov exponents or at least limits of stability of synchronized state  can be obtained analytically.
More detailed nature of synchronization transition is studied numerically.
We show that in the case of discontinuous maps, synchronization is also a discontinuous transition, despite continuously  vanishing (as a function of coupling strength) Lyapunov exponent(s).
For continuous maps this transition is found to be continuous.
We also study partial synchronization of nonidentical maps, and its relation with Lyapunov exponents.
We show that for tent maps at partial synchronization there are substantial changes in the probability distribution of the synchronization error.
But for Bernoulli maps partial synchronization is merely a smooth crossover between synchronized and nonsynchronized regimes.
Finally, we examine $d=4$ maps which might be considered as describing synchronization of two identical two-dimensional systems.
This case requires the most subtle analysis, since even in the synchronized state the system is relatively complicated (synchronized manifold is two-dimensional in this case).
Using some arguments, based on the spectral properties of Jacobian of this maps, we locate the point where the synchronized state becomes stable.
In some cases, this is most likely a global attractor since numerical calculations show that synchronization transition takes place also at this point.
However, in some cases, a basin of the synchronized attractor is riddled with basin of another chaotic attractor~\cite{RIDDLE}.
We suggests that such a coexistence of synchronized and nonsynchronized chaotic attractors might be a low-dimensional analog of the so-called stable chaos that was found in some spatially extended systems~\cite{LIVI}.
\section{Two-dimensional maps}
Objects of our main interest are linear maps.
First let us examine a two-dimensional map of the form
\begin{subequations}
\begin{eqnarray}
x_{n+1} & = & f((q-\epsilon)x_n+(1-q)y_n)\label{map2a}\\
y_{n+1} & = & f(qx_n+(1-q)y_n)
\label{map2b}
\end{eqnarray}
\end{subequations}
where $f(x)$ is a piecewise linear function, $q>0$, and $n=1,2\ldots$ can be interpreted as a time variable.
In the present paper we will use a generalized Bernoulli shift map
$f(x)=ax{\rm mod}(1)$ and a tent map
\begin{equation}
f(x)=\left\{ \begin{array}{ll}
ax & \textrm{for $0<x<1/a$} \\
-ax+2 & \textrm{for $1/a<x<1$}
\end{array}\right.
\label{tent}
\end{equation}
where $a\leq 2$ and in most of our numerical calculations we use $a=3/2$.
Certain aspects of dynamics of piecewise linear maps were already 
studied~\cite{LINEAR}.

First, let us consider the case of identical systems ($\epsilon=0$).
Since the Bernoulli map has a constant slope (except a set of zero measure),
Lyapunov exponents $\lambda_1$ and $\lambda_2$ of the this map 
can be obtained from the eigenvalues of the Jacobian of map (\ref{map2a})-(\ref{map2b})
\begin{equation}
\mathbf{T_2}=\left(\begin{array}{cc}
aq & a(1-q) \\
a(1-q) & aq 
\end{array}\right)
\label{matrix2}
\end{equation}
Lyapunov exponents of the tent map are  more difficult to obtain since 
in this case the Jacobian depends on the values of the arguments
of the maps in (\ref{map2a})-(\ref{map2b}).
In particular, Jacobian takes the form (\ref{matrix2}) only when both arguments are smaller than $1/a$.
When the argument of (\ref{map2a}) is smaller than $1/a$ but the argument of (\ref{map2b}) is larger, the Jacobian becomes
\begin{equation}
\mathbf{T_2'}=\left(\begin{array}{cc}
aq & a(1-q) \\
-a(1-q) & -aq 
\end{array}\right)
\label{matrix2prime}
\end{equation}
Two other possibilities lead to matrices $\mathbf{T_2''}=-\mathbf{T_2'}$ (larger, smaller) and $\mathbf{T_2'''}=-\mathbf{T_2}$ (both larger).
To determine Lyapunov exponents for tent map we would have to consider infinite product
of these four matrices in order determined by the dynamics 
(\ref{map2a})-(\ref{map2b}), which usually constitutes a formidable analytical task~\cite{RANDOMM}.
However, in our case the problem of calculating Lyapunov exponents simplifies greatly provided that the difference $x_n-y_n$ is negligibly small.
This is always the case in the synchronized phase and from the limits of stability of this phase we will be able to locate the synchronization transition.
When the difference $x_n-y_n$ is small the instances when one of the arguments is larger and the other smaller than $1/a$ are very rare.
One can thus neglect contributions coming from matrices 
$\mathbf{T_2'}$ and $\mathbf{T_2''}$.
Consequently, Lyapunov exponents in this case are obtained from the eigenvalues of the matrix $\mathbf{T_2}$ only.

Diagonalizing $\mathbf{T_2}$ one obtains the following Lyapunov exponents 
$\lambda_1={\rm ln}(a)$ with the eigenvector $\mathbf{e_1}=(1,1)$ and $\lambda_2={\rm ln}[a(2q-1)]$ with
$\mathbf{e_2}=(1,-1)$.
Let us notice, that $\mathbf{e_2}$ is a base vector of a one dimensional transversal manifold on which $x_n \neq y_n$.
Moreover, $\lambda_2=0$ at $q=\frac{a+1}{2a}$, that at the same time locates the synchronization transition of system (\ref{map2a})-(\ref{map2b}) both for Bernoulli and tent map.

To check our results, we introduce the synchronization error $w=\langle|x_n-y_n|\rangle$, where $\langle\ldots\rangle$ denotes an average over time and initial conditions.
Numerical results, obtained also for other maps described later on, were averaged over typically $10^3$ runs with different initial conditions.
Each run has a length $10^7$ steps with $10^6$ steps discarded as initial transient.
In some cases, where fluctuations were strong or relaxation time was very long we made runs of $10^8$ steps with $10^7$ steps discarded.

\begin{figure}
\centerline{\epsfxsize=9cm
\epsfbox{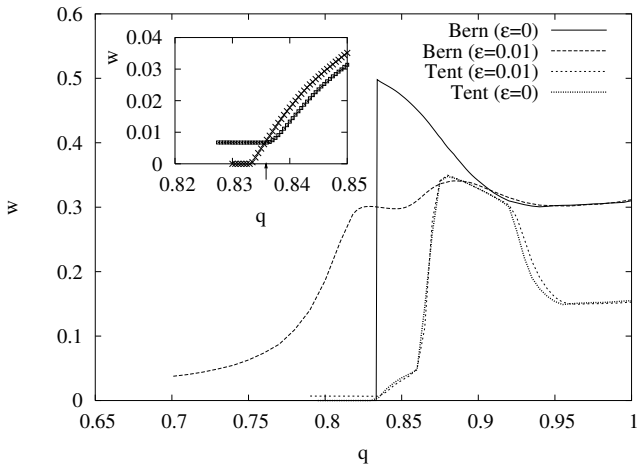}
}
%\figspace
\caption{
The synchronization error $w$ as a function of $q$ for the two-dimensional map (\ref{map2a})-(\ref{map2b}) with $a=3/2$.
Inset shows the behavior of $w$ for tent map in the vicinity of the transitions for $\epsilon=0$ ($\times$) and $\epsilon=0.01$ ($\Box$).
For $\epsilon=0.01$ $\lambda_2$ from Eq.~(\ref{lyap2}) vanishes at $q=q_c=0.83587\ldots$ that is indicated by a small arrow.
}
\label{steady2}
\end{figure}

Results of the numerical calculations for $a=3/2$ are shown in 
Fig.~\ref{steady2}.
They confirm that our system (\ref{map2a})-(\ref{map2b}) synchronizes at 
$q=\frac{a+1}{2a}=5/6(=0.833\ldots)$ both for Bernoulli and tent maps.
Let us notice, however, a difference in the behavior of these two maps.
Namely, while for tent map $w$ vanishes continuously to zero at $q=5/6$ it
has a discontinuous jump for Bernoulli map, although $\lambda_2$ vanishes continuously to zero in this case.

Now, let us consider the case of nonidentical systems, i.e., $\epsilon>0$.
In this case the Jacobian for the Bernoulli map has the form
\begin{equation}
\mathbf{T_2}=\left(\begin{array}{cc}
a(q-\epsilon) & a(1-q) \\
a(1-q) & aq 
\end{array}\right)
\label{matrix2epsilon}
\end{equation}
Its diagonalization gives the following Lyapunov exponents
\begin{equation}
\lambda_{1,2}={\rm ln}[\frac{a}{2}(2q-\epsilon\pm \sqrt{\epsilon^2+4(1-q)^2})]
\label{lyap2}
\end{equation}
For $a=3/2$ elementary calculations show that now \mbox{$\lambda_2<0$} for 
\begin{equation}
q<q_c=\frac{5-6\epsilon}{6-9\epsilon},
\label{where}
\end{equation}
As before $\lambda_1$ remains positive.

Since our systems are now nonidentical, the difference $x_n-y_n$ never vanishes and neglecting matrices $\mathbf{T_2'}$ and $\mathbf{T_2''}$ is
this time only an approximation.
But we expect that for small $\epsilon$ and in the partially synchronized state this approximation should be relatively accurate.
Thus, we assume that Eq.~(\ref{lyap2}) gives approximate Lyapunov exponents
at least inside partially synchronized phase.
Hence Eq.~(\ref{where}) locates the partial synchronization transition also for tent map.

Numerical calculations of $w$ as a function of $q$ are shown in 
Fig.~\ref{steady2}.
As expected, $w$ remains positive even for $q<q_c$ although for tent map there is a clear change of behavior around $q_c$.
For Bernoulli map $w$ seems to be a smooth function of $q$ even at $q_c$.
%%%%--------------------------------------
%%%---------------------------------------

\begin{figure}
\centerline{\epsfxsize=9cm
\epsfbox{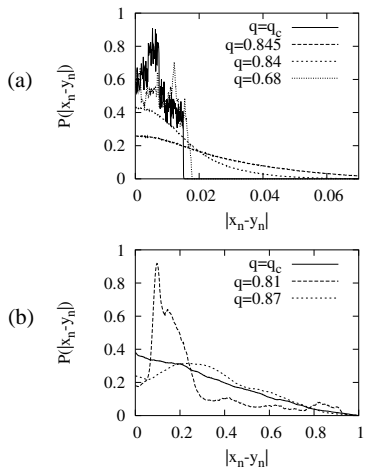}
}
%\figspace
\caption{
The nonnormalized probability distribution $P(|x_n-y_n|)$ for tent (a) and Bernoulli (b) map ($a=3/2$ and $q_c=0.83587\ldots$).
}
\label{binbertent}
\end{figure}

Recently, it was shown that partial synchronization in some continuous dynamical systems manifests through changes in the probability distribution
of the phase difference~\cite{ZHOU}.
To check whether this is more general property we calculated probability distribution of the synchronization error $P(|x_n-y_n|)$ and the results are shown in Fig.~\ref{binbertent}.
One can see that in the nonsynchronized phase of both tent and Bernoulli map 
$P(|x_n-y_n|)$ has a rather broad and smooth distribution.
However, for tent map at $q=q_c$ this distribution becomes very complex.
This complexity persists also in the synchronized phase ($q<q_c$).
On the other hand, there is no noticeable difference at $q=q_c$ for the Bernoulli map, although it develops a pronounced peak in the synchronized phase.
Taking into account also the smooth behavior of $w$ at $q=q_c$ it is legitimate to ask whether partial synchronization in the case of Bernoulli map  is a sharp and well defined transition or it is rather a smooth crossover.
In this context it would be desirable to examine a map that continuously interpolates between these two maps as for example a skewed tent map.
%%%%--------------------------------------
%%%---------------------------------------
\section{Three-dimensional maps}
As a next example let us consider a three-dimensional map of the form
\begin{subequations}
\begin{eqnarray}
x_{n+1} & = & f((q-\epsilon)x_n+0.5(1-q)(y_n+z_n))\label{map3a}\\
y_{n+1} & = & f(qy_n+0.5(1-q)(y_n+x_n))\label{map3b}\\
z_{n+1} & = & f(qz_n+0.5(1-q)(x_n+y_n))\label{map3c}
\end{eqnarray}
\end{subequations}
As before, Lyapunov exponents for the Bernoulli map are obtained from the eigenvalues of the following matrix
\begin{equation}
\mathbf{T_3}=\left(\begin{array}{ccc}
a(q-\epsilon) & a(1-q)/2 & a(1-q)/2\\
a(1-q)/2 & aq & a(1-q)/2\\ 
a(1-q)/2 & a(1-q)/2 & aq
\end{array}\right)
\label{matrix3}
\end{equation}
And similarly to the $d=2$ case, for $\epsilon=0$ and in the 
synchronized phase, matrix $\mathbf{T_3}$ contains also Lyapunov 
exponents of tent map (in the synchronized state all three variables are the same).

\begin{figure}
\centerline{\epsfxsize=8cm
\epsfbox{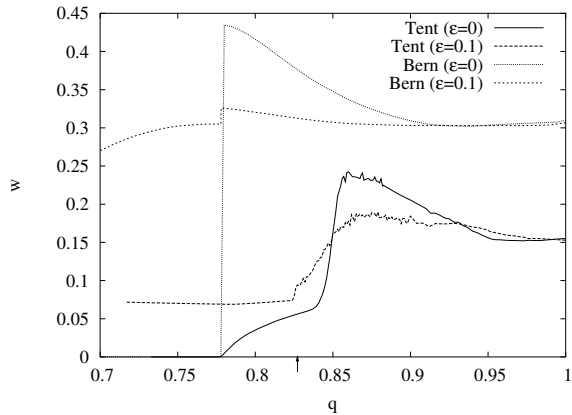}
}
%\figspace
\caption{
The synchronization error $w$ as a function of $q$ for the three-dimensional map (\ref{map3a})-(\ref{map3c}).
For $\epsilon=0.1$ $\lambda_3$ vanishes at $q=q_c=0.82717\ldots$ indicated by an arrow ($a=3/2$).
}
\label{tent3}
\end{figure}

For $\epsilon=0$ we obtain that the largest exponent $\lambda_1$ is always positive.
Two other exponents are degenerate $\lambda_2=\lambda_3$, their eigenvectors are the base of a two-dimensional transversal manifold, and for $a=3/2$ they vanish at $q=7/9(=0.777\ldots)$.
Numerical calculation of the synchronization error show (see 
Fig.~\ref{tent3}) that at this point, as expected, synchronization transition takes place.
Similarly to $d=2$ map, for Bernoulli map this transition is discontinuous.

For nonidentical systems ($\epsilon=0.1$) we find that $\lambda_3$ vanishes at $q=q_c=\frac{7-3\epsilon}{9(1-\epsilon)}=0.82717\ldots$.
On similar grounds as in the $d=2$ case, we consider it also as an approximate location of the partial synchronization transition in the tent map.
Figure~\ref{tent3} confirms this argument since one can see that  synchronization error $w$ for tent map indeed changes its behavior very close to this point.
For Bernoulli map $w$ seems to be again a smooth function around $q_c$.
But there is yet another transversal exponent, namely $\lambda_2$ (with the eigenvector ${\mathbf e_2}$=(0,1,-1)) that vanishes as in the $\epsilon=0$ case i.e., at $q=7/9$.
At this point complete synchronization of $y$ and $z$ systems takes place (they are identical).
At this point $w$ has a jump but only for Bernoulli map.

We also examined the probability distribution $P(x_n-y_n)$.
Our results (not shown) indicate that for tent map $P(|x_n-y_n|)$ has a complex structure in the synchronized phase ($q<q_c$).
On the other hand no appreciable change is seen for Bernoulli map which is similar to the $d=2$ case that we already discussed.
It suggests that for discontinuous maps there are almost no indications of a qualitative change at partial synchronization (except of the vanishing Lyapunov exponent) and most likely it is only a smooth crossover.
%%%%--------------------------------------
%%%---------------------------------------
\section{Four-dimensional maps}
As a final example let us consider a four-dimensional map of the form
\begin{subequations}
\begin{eqnarray}
x_{n+1} & = & f(0.9x_n+qy_n+0.1z_n)\label{map4a}\\
y_{n+1} & = & f(0.9y_n+qx_n+0.1u_n)\label{map4b}\\
z_{n+1} & = & f(0.9z_n+qu_n+0.1x_n)\label{map4c}\\
u_{n+1} & = & f(0.9u_n+qz_n+0.1y_n)\label{map4d}
\end{eqnarray}
\end{subequations}
This map describes a coupling of four identical systems $x$, $y$, $z$, and $u$.
Increasing $q$ increases the coupling $x-y$ and $z-u$ and for sufficiently large $q$ the system enters a synchronized state were $x_n=y_n$ and 
$z_n=u_n$.
Numerical coefficients in this map are chosen in such a way that subsystems 
$x-z$ and $y-u$ remain nonsynchronized and chaotic.

\begin{figure}
\centerline{\epsfxsize=9cm
\epsfbox{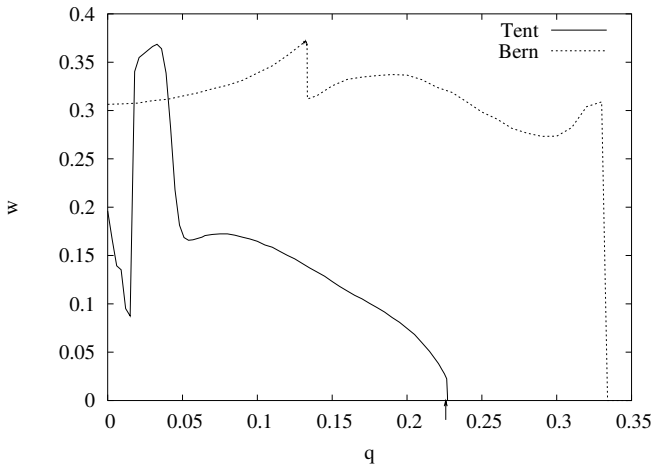}
}
%\figspace
\caption{
The synchronization error $w$ as a function of $q$ for the four-dimensional map (\ref{map4a})-(\ref{map4d}).
We also measured the difference $\langle|z_n-u_n|\rangle$ and found that within statistical errors, as expected, it equals to $w$.
An arrow indicates a value $q=0.225875\ldots$.
}
\label{tent4}
\end{figure}

Lyapunov exponents for the Bernoulli map are obtained from the eigenvalues of the following matrix
\begin{equation}
\mathbf{T_4}=\left(\begin{array}{cccc}
0.9a & aq & 0.1a & 0\\
aq & 0.9a & 0 & 0.1a\\
0.1a & 0 & 0.9a & aq\\
0 & 0.1a & aq & 0.9a
\end{array}\right)
\label{matrix4}
\end{equation}
Diagonalizing $\mathbf{T_4}$ we find that two Lyapunov exponents 
$\lambda_1={\rm log}[a(1+q)]$ and $\lambda_2={\rm log}[a(0.8+q)]$ are always positive.
Corresponding eigenvectors $\mathbf{e_1}=(1,1,1,1)$ and 
$\mathbf{e_2}=(1,1,-1,-1)$ are the base of the synchronized subspace (on which $x_n=y_n$ and $z_n=u_n$).
There are also two transversal exponents $\lambda_3={\rm log}[a(1-q)]$ and 
$\lambda_4={\rm log}[a(0.8-q)]$ with eigenvectors 
$\mathbf{e_3}=(-1,1,-1,1)$ and $\mathbf{e_4}=(1,-1,-1,1)$.
For $a=3/2$ one can see that $\lambda_3=0$ at $q=1/3$ and $\lambda_4=0$ at $q=2/15(=0.1333\ldots)$.
One can thus expect that the system synchronizes when the largest transversal exponent vanishes i.e., at $q=1/3$.
Indeed, such a behavior is seen for the Bernoulli map (Fig.~\ref{tent4}).
In this case one can also see a jump in $w$ at $q=2/15$ that is a trace of  vanishing of $\lambda_4$.
However, for tent map a different behavior is seen, and synchronization takes place around $q=0.23$.

To explain such a behavior we have to notice that the synchronized manifold is in this case two dimensional and in the synchronized state the system is described by two variables e.g., $x_n(=y_n)$ and $z_n(=u_n)$.
It means that in the synchronized state the difference $x_n-z_n$ does not vanish.
Consequently, arguments of maps (\ref{map4a})-(\ref{map4d}) might differ and we have to take into account that in some cases Jacobian of the map has a different form than $\mathbf{T_4}$.
For example, when arguments of the first two maps (\ref{map4a})-(\ref{map4b}) are both smaller than $1/a$ and arguments of maps (\ref{map4c})-(\ref{map4d}) are both larger than $1/a$, then the Jacobian is given by the following matrix
\begin{equation}
\mathbf{T_4'}=\left(\begin{array}{cccc}
0.9a & aq & 0.1a & 0\\
aq & 0.9a & 0 & 0.1a\\
-0.1a & 0 & -0.9a & -aq\\
0 & -0.1a & -aq & -0.9a
\end{array}\right)
\label{matrix4prime}
\end{equation}
In the following, similarly to the $d=2$ and 3 cases, we will analyse the stability of the synchronized state.
When differences $x_n-y_n$ and $z_n-u_n$ are negligibly small then arguments of maps (\ref{map4a}) and (\ref{map4b}) are almost the same, and  arguments of maps (\ref{map4c}) and (\ref{map4d}) are also almost the same.
Thus, the situation is analogous to the $d=2$ example and there are only four possible forms of the Jacobian given by the matrices $\mathbf{T_4}$,
$\mathbf{T_4'}$, $\mathbf{T_4''}=-\mathbf{T_4'}$, and 
$\mathbf{T_4'''}=-\mathbf{T_4}$.
However, since $\mathbf{T_4}$ and $\mathbf{T_4'''}$ do not commute with 
$\mathbf{T_4'}$ and $\mathbf{T_4''}$ finding Lyapunov exponents
of our map even under such a simplifying assumption is very difficult.
But a closer look at the spectral properties of these matrices reveals certain interesting properties (with respect to their spectrum, it is of course sufficient to examine for example only matrices $\mathbf{T_4}$ and 
$\mathbf{T_4'}$).
Calculating eigenvalues of $\mathbf{T_4'}$ one finds the following Lyapunov exponents:
$\lambda_1'=\lambda_2'={\rm ln}[a\sqrt{(1+q)(0.8+q)}]$, 
$\lambda_3'=\lambda_4'={\rm ln}[a\sqrt{(1-q)(0.8-q)}]$.
These Lyapunov exponents are associated only with matrix $\mathbf{T_4'}$ and they are not Lyapunov exponents  of map (\ref{map4a})-(\ref{map4d}).
Exponents $\lambda_1'$ and $\lambda_2'$ are positive and their eigenvectors are within the synchronized manifold.
Eigenvectors of exponents $\lambda_3'$ and $\lambda_4'$ are within the transversal manifold.
These exponents change sign at 
$q=q_c=0.9-\sqrt{0.01+a^{-2}}$, that for $a=3/2$ equals $0.225875\ldots$, i.e., the value that is very close to the actual transition in the tent map as obtained numerically (Fig~\ref{tent4}).
However, for such a value of $q$ only one of the transversal exponents of 
$\mathbf{T_4}$ is negative ($\lambda_4$) and the other ($\lambda_3$) remains positive.
But for $q=0.9-\sqrt{0.01+a^{-2}}$ and interesting property of matrix 
$\mathbf{T_4}$ holds namely $\lambda_3+\lambda_4=0$.
It suggests that indeed, at this value a synchronization transition of the tent map might take place, however a closer look into the dynamics of  
our map (\ref{map4a})-(\ref{map4d}) is needed.
%%%%--------------------------------------
%%%---------------------------------------

\begin{figure}
\centerline{\epsfxsize=9cm
\epsfbox{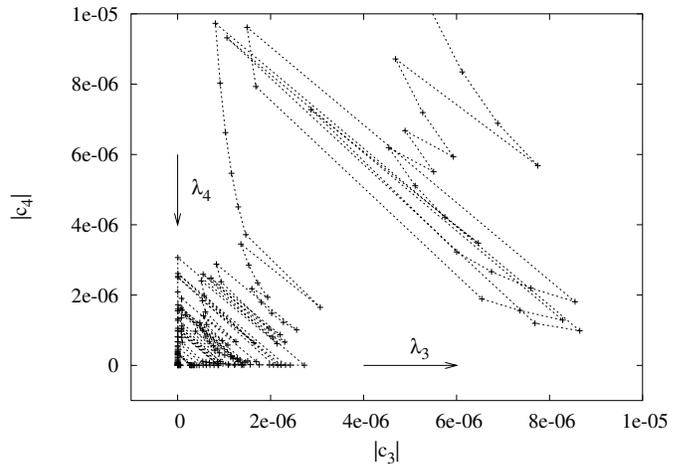}
}
%\figspace
\caption{
A typical trajectory in a transversal subspace $(|c_3|,|c_4|)$ for 
$q=0.25$ and $a=3/2$ (only absolute values of coordinates $c_3,c_4$ are plotted).
Arrows indicate the action of Lyapunov exponents of $\mathbf{T_4}$.
Due to swapping of coordinates (see text) the system evolves toward the synchronized state (0,0).
}
\label{saddle}
\end{figure}

Let us analyse the dynamics of our system for $q>0.9-\sqrt{0.01+a^{-2}}$ as projected on the transversal manifold.
Each state vector $(x_n,y_n,z_n,u_n)$ is decomposed in the base of
eigenvectors of $\mathbf{T_4}$ as
$(x_n,y_n,z_n,u_n)=c_1\mathbf{e_1}+c_2\mathbf{e_2}+c_3\mathbf{e_3}+
c_4\mathbf{e_4}$, and then we look at the evolution of transversal coordinates $c_3$  and $c_4$.
A typical trajectory is shown in Fig.~\ref{saddle}.
One can see that after an initial transient, the system locates
essentially on the axes of eigenvectors and gradually approaches the point (0,0) i.e., a synchronized state, swapping from time to time its coordinates.
Actually, such a behavior can be deduced from Eqs.~(\ref{map4a})-(\ref{map4d}).
First, we express coordinates in the transversal subspace as 
$c_3=(-x_n+y_n-z_n+u_n)/4=(-\delta_{1}-\delta_{2})/4$ and 
$c_4=(x_n-y_n-z_n+u_n)/4=(\delta_{1}-\delta_{2})/4$, where 
$\delta_1=x_n-y_n$ and $\delta_2=z_n-u_n$.
Since  $\delta_1,\delta_2$ are small it implies that the system evolves 'in pairs'.
Namely, either both arguments of maps (\ref{map4a})-(\ref{map4b}) are smaller or both are larger than 
$1/a$ and either the first or the second definition of the tent map (\ref{tent}) is applied to this pair.
The same applies for the second pair of maps (\ref{map4c})-(\ref{map4d}).
The swap of coordinates appears when two pairs evolve according to different rules.
Indeed, let us assume for example that the first pair of arguments is smaller and the second pair is larger than $1/a$.
Using the evolution equations (\ref{map4a})-(\ref{map4d}) we find that
initial coordinates ($(-\delta_{1}-\delta_{2})/4,(\delta_{1}-\delta_{2})/4$) are transformed into ($-a(0.8-q)(\delta_{1}-\delta_{2})/4,
a(1-q)(\delta_{1}+\delta_{2})/4$).
Up to the sign and the multiplicative factor of the order of unity this is indeed the swap of coordinates.
Since for $q>0.9-\sqrt{0.01+a^{-2}}$  the sum of transversal exponents 
$\lambda_3+\lambda_4<0$, shrinking along $\mathbf{c_4}$ axis prevails 
over repelling along $\mathbf{c_3}$ axis.
Without swapping the system would drift to infinity along $\mathbf{c_3}$ axis.
But swapping turns the system back almost on the $\mathbf{c_4}$ axis where it is
subjected (mainly) to shrinking.
Of course, swapping can also put the system again on the $\mathbf{c_3}$ axis where it is repelled, but sooner or later it will return on the 
$\mathbf{c_4}$ axis, where it will be again subjected to shrinking.
Swapping of coordinates is actually described by  matrices $\mathbf{T_4'}$ or $\mathbf{T_4''}$.
For $q>0.9-\sqrt{0.01+a^{-2}}$ these matrices have negative transversal 
exponents and their action in the transversal manifold can only shrink the system toward the origin (0,0).
As a result of this combined action (swapping and shrinking faster than repelling), the system must end up in the synchronized state.
On the other hand for $q<0.9-\sqrt{0.01+a^{-2}}$ both mechanisms which attracted the system toward the synchronized state fail.
Repelling along $\mathbf{c_3}$ axis prevails over shrinking along
along $\mathbf{c_4}$ axis and transversal exponents of $\mathbf{T_4'}$ and
$\mathbf{T_4''}$ are positive.
As a result the system remains in the nonsynchronized state.
The value $q=0.9-\sqrt{0.01+a^{-2}}=0.225875\ldots$ (for $a=3/2$) is indicated in Fig.~\ref{tent4} with an arrow and is in a very good agreement with numerical calculations.
%%%%--------------------------------------
%%%---------------------------------------

\begin{figure}
\centerline{\epsfxsize=9cm
\epsfbox{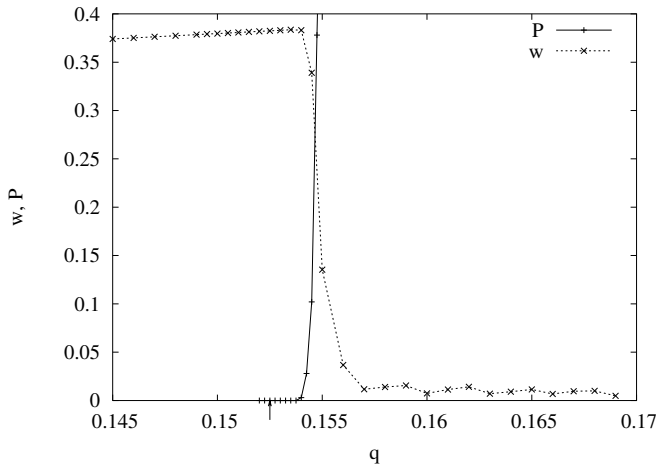}
}
%\figspace
\caption{
The synchronization error $w$ and the probability $P$ of reaching the synchronized state as a function of $q$ for the four-dimensional map (\ref{map4a})-(\ref{map4d}) with $a=1.35$.
Arrow indicates a value $q=0.1525397\ldots$, where a synchronized state becomes stable.
}
\label{tent4a135}
\end{figure}

There is, however, a silent assumption in the above considerations, namely that on the way to the synchronized state the system will not fall into the basin of another attractor that might possibly exist.
Actually, it is already known that 
basins of synchronized chaotic attractors are sometimes riddled with basins of some other attractors, even when transversal Lyapunov exponents are negative~\cite{RIDDLE}.
As a result, only some initial conditions will lead the system to the synchronized state, and the other will lead to the other nonsynchronized attractor.
And indeed, we found that for $a=1.35$ such a scenario most likely takes place (Fig.~\ref{tent4a135}).
For $a=1.35$ we obtain that the synchronized state becomes stable  for $q>0.9-\sqrt{0.01+a^{-2}}=0.1525397\ldots$.
Although the synchronization error $w$ does not vanish at this point,
we observed that for $q\gtrsim 0.154$ there is a finite fraction $P$ of initial conditions that leads relatively fast to the synchronized state.
Remaining initial conditions lead the system to another chaotic attractor.
As a result, $w$ that is an average over runs with different initial conditions, remains positive.
%%%%--------------------------------------
%%%---------------------------------------

\begin{figure}
\vspace{-2cm}
\centerline{\epsfxsize=9cm
\epsfbox{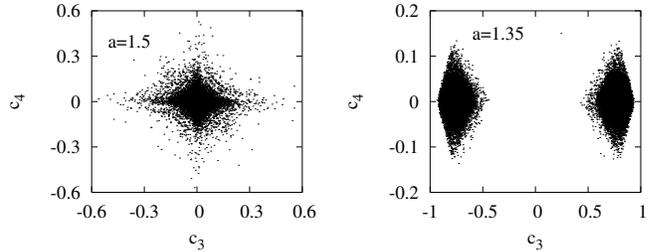}
}
%\figspace
\caption{
Critical point attractors as projected on the transversal manifold 
$(c_3,c_4)$.
}
\label{attractor}
\end{figure}

To show that $a=1.35$ and $a=1.5$ cases are qualitatively different we plot attractors at their critical points $q=q_c=0.9-\sqrt{0.01+a^{-2}}$ as projected on the transversal manifold $(c_3,c_4)$ (Fig.~\ref{attractor}).
These plots are obtained from several independent runs after discarding $10^8$ steps and recording next $3\cdot 10^4$ steps.
We noticed, that when $q$ is smaller than $q_c$ no appreciable change is seen in the structure of these attractors.
However, when $q$ exceeds even slightly the critical value important differences takes place.
Namely for $a=1.5$ the attractor shrinks to the origin (0,0).
For $a=1.35$ the origin also becomes an attractor but the nonsynchronized attractors persist as well, with only small changes.
In Fig.~\ref{tent4a135} probability $P$ of reaching a synchronized state seems to be positive only for $q\gtrsim 0.154$, which might be a consequence of the numerical method.
Namely we considered that a system reached a synchronized state if after $10^6$ steps its coordinates satisfy $|x_n-y_n|+|z_n-u_n|<10^ {-5}$.
But close to the critical poind $q_c$ an approach to the synchronized state is very slow and we expect that when larger number of iteration steps were used, we would find positive $P$ for any $q>q_c$.
Let us also notice that the fact that for $a=1.5$ at the critical point 
the attractor is different than the origin implies a finite value of synchronization error $w$ at this point i.e., the synchronization transition is most likely discontinuous in this case.

Finally, let us notice that in synchronization of some spatially extended systems, like Coupled-Map Lattices, it was observed that in presence of negative transversal Lyapunov exponents nonsynchronized state might 
exist~\cite{AHLERS}.
A similar behavior appears in some other extended systems and is known as stable chaos~\cite{LIVI}.
In our opinion, the behavior of our system for $a=1.35$ bears some similarity to this effect.
Although we did not calculate the transversal Lyapunov exponents explicitly, it seems very likely that for $q>q_c$ they are negative (since the synchronized state is stable, and that is also confirmed numerically because a finite fraction of initial conditions leads to this state).
Nevertheless, a finite fraction of initial conditions lead the system to a nonsynchronized state.
An important difference is that in extended systems exhibiting stable chaos a nonsynchronized state is a generic state of the dynamics, that is reached from almost any initial condition.
Another property of such extended systems is that a nonsynchronized state after a certain threshold looses its stability but no such a threshold was found in our 4-dimensional map.
But it might be that these features are consequences of high dimensionality of dynamics of an extended system.
%%%%--------------------------------------
%%%---------------------------------------
\section{Conclusions}
In conclusion, we studied synchronization of low-dimensional linear chaotic maps.
Even for such simple maps finding Lyapunov exponents might be very difficult.
But as we have shown, the problem significantly simplifies in the synchronized phase which enabled us to locate exactly the  synchronization transitions.
Our results show qualitative differences between the behavior of Bernoulli and tent map
concerning the type of the synchronization transition (discontinuous for Bernoulli map, and continuous for tent map), or even the location of the transition (in the $d=4$ case).
Morever, very different behavior is seen when maps are nonidentical and only partial synchronization takes place.
It might be that it is a discontinuity of the Bernoulli map that is responsible for some of these differences.
Noting the similarity to the behavior of some extended systems, it would be interesting to examine whether the origin of the stable chaos indeed can be relateded with riddled basins of attractors in low-dimensional systems like those examined in the present work.

This work was partially supported by the Swiss National Science Foundation
and the project OFES 00-0578 "COSYC OF SENS".
Some of our calculations were done on 'openMosix Cluster' built and
administrated by Lech D\c{e}bski.
%%%%%%%%%%%%%%%%%%%%%%%%%%%%%%%%%%%%%%%%%%%%%%%%%%%%%%%%%%%%%%%%%%%%%%%%%%%%%%
%%%%%%%%%%%%%%%%%%%%%%%%%%%%%%%%%%%%%%%%%%%%%%%%%%%%%%%%%%%%%%%%%%%%%%%%%%%%%%%

%%%%%%%%%%%%%%%%%%%%%%%%%%%%%%%%%%%%%%%%%%%%%%%%%%%%%%%%%%%%%%%%%%%%%%%%%%%%%%%
%%%%%%%%%%%%%%%%%%%%%%%%%%%%%%%%%%%%%%%%%%%%%%%%%%%%%%%%%%%%%%%%%%%%%%%%%%%%%%%
\end {document}